\documentclass[prc,preprint]{revtex4}

\usepackage{graphicx}

\begin{document}

\title{{Calculation of the Structure Properties of a Strange Quark Star in the Presence
of Strong Magnetic Field Using a Density Dependent Bag Constant }}

\author{{ Gholam Hossein Bordbar$^{1,2}$
\footnote{Corresponding author. E-mail:
bordbar@physics.susc.ac.ir}}, { Hajar Bahri$^{1}$} and { Fatemeh Kayanikhoo$^{1}$}}
\affiliation{$^{1}$Department of Physics,
Shiraz University, Shiraz 71454, Iran\\
and \\ $^{2}$Research Institute for Astronomy and Astrophysics of Maragha,
P.O. Box 55134-441, Maragha, Iran}

%%%%%%%%%%%%%%%%%%%%%%%%%%%%%%%%%%%%%%%%%%%%%%%%%%%%%%%%%%%%%%%%%%%%%%%%%%%%%%%%%%%%%%

\begin{abstract}
In this article we have calculated the structure properties  of
a strange quark star in static model in the presence of a strong magnetic field using MIT bag model
with a density dependent bag constant. To parameterize the density dependence of
bag constant, we have used our results for the lowest order constrained
variational  calculation of the asymmetric nuclear matter. By calculating the equation of state of
strange quark matter, we have shown that the pressure of this system increases by increasing both density and
magnetic field.  Finally, we have investigated the effect of density dependence of bag constant
on the structure properties  of  strange quark star.
\end{abstract}
\maketitle
%
%%%%%%%%%%%%%%%%%%%%%%%%%%%%%%%%%%%%%%%%%%%%%%%%%%%%%%%%%%%%%%%%%%%%%%%%%%%%%%%%%%%%%
\section{Introduction}
The core of a neutron star has been formed from nuclear matter, composed of the neutrons, protons, electrons
(for assurance negation of electric charge) and other particles like pions, mesons and etc
(Lattimer \& Prakash~\cite{rk1}).
It is known that nuclear matter is meta-stable, that after releasing a lot of
energy converts into strange quark matter (SQM)
to achieve stability. This quark matter is the most stable state of matter that has been
known until now. Thus, there is a new class of compact stars that come from the collapse
of neutron stars, and are more stable compared to the neutron stars (Farhi \& Jaffe~\cite{rk2}).
The best candidates for this conversion are the neutron stars with masses of $1.5-1.8\ M_{\odot}$ and quick spin
(Drake et al.~\cite{rk3}; Li et al.~\cite{rk4}; Weber~\cite{rk5}).

The collapse of a neutron star may lead to a strange quark star (SQS) or a hybrid star.
Also under special conditions, an SQS may be directly born from the core collapse of a type II supernova.
An SQS, from its center to surface is made from SQM,
and on its surface may exist a nuclear layer (Glendenning \& Weber~\cite{rk6}).
Hybrid stars are the ones with  cores composed of SQM (Bhattacharyya et al.~\cite{rk7}).
Here, we just consider the structure properties of SQS.

The mass and density of an SQS is between the mass and density
of a neutron star and that of a black hole. The mass-radius relation for an SQS is as $M \propto R^{3}$
which is different from that of a neutron star.
This star does not have the minimum mass. For an SQS  with $1M_{\odot} \leq M \leq 2M_{\odot}$, the radius is about $10\ km$
(Farhi \& Olinto~\cite{rk8}; Shapiro \& Tenkolsky~\cite{rk9}).

Recent observations indicate that the object SWIFT J1749.4-2807 may be an SQS (Yu \& Xu~\cite{rk17}).
The given results by Chandra observations also show that the objects RX J185635-3754 and 3C58 may be bare strange stars
(Prakash et al.~\cite{rk55}).
It is known that the compact objects such as neutron star, pulsars, magnetars and strange quark stars, are under the
influence of strong magnetic fields which are typically about $10^{15} - 10^{19}\ G$
(Kouveliotou et al.~\cite{rk10}, \cite{rk11}; Haensel et al.~\cite{rk12}; Glendenning~\cite{rk13};
Weber~\cite{rk14}; Camenzind~\cite{rk15}).
Therefore, in astrophysics, it is of special interest to study the effect of a strong magnetic field on the properties of SQM.
We note that in the presence of a magnetic field, the conversion of neutron stars
to bare quark stars cannot take place unless the value of the magnetic field exceeds $10^{20}\ G$
(Ghosh \& Chakrabarty~\cite{rk16}).

In recent years, we have calculated the maximum gravitational mass and other structure properties
of a neutron star with a quark core at zero (Bordbar et al.~\cite{rk53}) and finite temperatures
(Yazdizadeh \& Bordbar~\cite{rk54}).
We have also computed the structure properties of  SQS at zero temperature
(Bordbar et al.~\cite{rk18}) and finite temperature (Bordbar et al.~\cite{rk19}).
We have also calculated the structure of a magnetized SQS using MIT bag model with a fixed bag constant
($90\ \frac{MeV}{fm^{3}}$)
(Bordbar \& Peyvand~\cite{rk20}). In the present work, we investigate the effect of density dependence of
bag constant on the structure of an SQS in the presence of strong magnetic field.
 %-------------------------------------------------------------------------------------
%-------------------------------------------------------------------------------------
\section{Computation of strange quark matter equation of state in the presence of magnetic field}
\label{I}

The equation of state (EOS) of strange quark matter (SQM) plays an important role for determining the structure
of stars at high densities. To obtain EOS of SQM, there are different models based
on Quantum Chromodynamics (QCD). At present, it isn't possible to achieve an exact EOS of SQM
by primary principles of QCD. Therefore, scientists have tried to find approximate methods by combining
the basic features of QCD, for example, MIT bag model
(Chodos et al.~\cite{rk21}; Weber~\cite{rk14}; Peshier et al.~\cite{rk45}; Alford et al.~\cite{rk46}),
NJL model (Rehberg et al.~\cite{rk47}; Hanauske et al.~\cite{rk48}; R¨uster \& Rischke~\cite{rk49};
Menezes et al.~\cite{rk50}), and perturbative QCD model
(Baluni~\cite{rk51}; Fraga et al.~\cite{rk52}; Farhi \& Jaffe~\cite{rk2}).

In MIT bag model, the quarks in the bag are considered as a free Fermi gas, and the energy per volume for SQM is equal to the kinetic
energy of the free quarks plus a bag constant ($ \mathcal{B} $) (Chodos et al.~\cite{rk21}).
The bag constant $ \mathcal{B} $ can be
interpreted as the difference between the energy densities of the noninteracting quarks and the
interacting ones. Dynamically, its role is as the pressure that keeps the quark gas in constant density
and potential. In the initial MIT bag
model, different values such as $55$ and $90\ \frac{MeV}{fm^{3}}$ are considered for the bag constant.
As we know, the density of SQM increases from surface to the core
of SQS, therefore  using a density dependent bag constant instead of a fixed
bag constant is more suitable.

\subsection{Density dependent bag constant}
The analysis of the experimental data achieved at CERN shows that
the quark-hadron transition happens at a density about seven times
the normal nuclear matter energy density ($156\ MeV fm^{-3}$)
(Heinz~\cite{rk22}; Heinz \& Jacobs~\cite{rk44};  Farhi \&
Jaffe~\cite{rk2}).
However theoretically, for no density-independent value of bag
constant the hadron to quark matter transition takes place (Burgio
et al.~\cite{rk26}). Therefore, it is essential to use a density
dependent bag constant.
Recently, a density dependent form has been also considered for $
\mathcal{B} $ (Adami \& Brown~\cite{rk23}; Jin \&
Jenning~\cite{rk24}; Blaschke et al.~\cite{rk25}; Burgio et
al.~\cite{rk26}). The density dependence of $ \mathcal{B} $ is
highly model dependent.
According to the hypothesis of a constant energy density along the
transition line, Burgio et al. tried to determine a range of
possible values for B by exploiting the experimental data obtained
at the CERN SPS (Burgio et al.~\cite{rk26}). By assumption that the
transition to quark-gluon plasma is determined by the value of the
energy density only, they estimated the value of bag constant and
its possible density dependence. They attempted to provide effective
parameterizations for this density dependence, trying to cover a
wide range by considering some extreme choices in such a way that at
asymptotic densities, the bag constant has some finite value.
They employed a Gaussian form as follows
\begin{equation}\label{1}
 \mathcal{B}(\rho) = \mathcal{B}_{\infty} + ( \mathcal{B}_{0} -
  \mathcal{B}_{\infty})e^{- \gamma(\rho/\rho_{0})^{2}} .
\end{equation}
The parameter $\mathcal{B}_{0} = \mathcal{B}(\rho = 0)$ is constant and equal to
$\mathcal{B}_{0} = 400\ \frac{MeV}{fm^{3}}$.
In the above equation, $\gamma$ is a numerical parameter which is usually equal to $\rho_{0} \approx  0.17\ fm^{-3}$,
the normal nuclear matter density. $ \mathcal{B}_{\infty}$ depends only on the free parameter $ \mathcal{B}_{0} $.

The value of the bag constant ($ \mathcal{B} $) should be compatible with experimental data.
The experimental results at CERN-SPS confirms a proton fraction $x_{pt} = 0.4$
(Heinz~\cite{rk22}; Heinz \& Jacobs~\cite{rk44}; Burgio et al.~\cite{rk26}).
Therefore, we use the equation of state of asymmetric nuclear matter to evaluate $\mathcal{B}_{\infty}$.
We use the lowest order constrained variational (LOCV) many-body method based on the cluster
expansion of the energy for calculating the equation of state of asymmetric nuclear matter as follows
(Bodbar \& Modarres~\cite{rk27}, \cite{rk28}; Modarres \& Bordbar~\cite{rk29}; Bordbar \&
Bigdeli~\cite{rk30}, \cite{rk31}, \cite{rk32}, \cite{rk33}; Bigdeli et al.~\cite{rk34}; Bigdeli et al.~\cite{rk35}).

The asymmetric nuclear matter is defined as a system consisting of $Z$ protons
($pt$) and $N$ neutrons ($nt$) with the total number density $\rho = \rho_{pt} + \rho_{nt}$ and proton fraction
$x_{pt} = \frac{\rho_{pt}}{\rho}$, where $\rho_{pt}$ and $\rho_{nt}$ are the number densities of protons and neutrons, respectively. For
this system, we consider a trial wave function of the form,
\begin{equation}\label{2}
\psi = F \phi,
\end{equation}
where $\phi$ is the slater determinant of the single-particle wave functions, and $F$ is the A-body correlation
operator ($A = Z + N$) which is given by
\begin{equation}\label{3}
F = \mathcal{S}\prod_{i>j}f(ij).
\end{equation}
In the above equation,  $ \mathcal{S} $ is a symmetrizing operator.

For the asymmetric nuclear matter, the energy per
nucleon up to the two-body term in the cluster expansion is as follows
\begin{equation}\label{4}
E([f]) = \frac{1}{A} \frac{< \psi | H | \psi >}{< \psi | \psi >} =
E_{1} + E_{2}.
\end{equation}
The one-body energy, $E_{1}$, is
\begin{equation}\label{5}
E_{1} = \sum_{i=1}^{2} \sum_{k_{i}} \frac{\hbar^{2} k_{i}^{2}}{2m},
\end{equation}
where labels 1 and 2 are used for proton and neutron respectively, and $k_{i}$ is the momentum
of particle $i$. The two-body energy, $E_{2}$, is given by
\begin{equation}\label{6}
E_{2} = \frac{1}{2A}\sum_{ij}< ij | \mathcal{V}(12) | ij - ji>,
\end{equation}
where
\begin{equation}\label{7}
\mathcal{V}(12) = - \frac{\hbar^{2}}{2m}[f(12),[\nabla_{12}^{2},f(12)] ]
+ f(12)V(12)f(12).
\end{equation}
In Eq. (\ref{7}), $f(12)$ and $V(12)$ are the two-body correlation and nucleon-nucleon
potential, respectively. In our calculations, we use $UV_{14} + TNI$ nucleon-nucleon potential
(Lagaris \& Pandharipande~\cite{rk36}, \cite{rk37}). We minimize the two-body energy with respect
to the variations in the correlation functions subject to the normalization constraint. From the
minimization of the two-body energy, we get a set of differential equations. By numerically solving these
differential equations, we can calculate the correlation functions. The two-body energy is obtained using
these correlation functions and then we can calculate the energy of asymmetric nuclear matter.
The procedure of these calculations has been fully discussed in reference (Bordbar \& Modarres~\cite{rk28}).

The experimental results at CERN-SPS confirm a proton
fraction $x_{pt} = 0.4$ (Heinz~\cite{rk22}; Heinz \& Jacobs~\cite{rk44}; Burgio et al.~\cite{rk26}).
Therefore, to calculate $\mathcal{B}_{\infty}$, we use our results of the above formalism for the
asymmetric nuclear matter characterized by a proton fraction $x_{pt} = 0.4$.
By assuming that the hadron-quark transition takes place at the energy
density equal to $1100\ MeV fm^{-3}$ (Heinz~\cite{rk22}; Burgio et al.~\cite{rk26}), we find that the
baryonic density of nuclear matter corresponding to this value of the energy
density is $\rho_{B} = 0.98\ fm^{-3}$ (transition density). At densities
lower than this value, the energy density of SQM is higher than that of the nuclear matter.
With increasing the baryonic density, these two energy densities become equal at the transition density,
and above this value, the nuclear matter energy density remains always higher.
Later, we determine $\mathcal{B}_{\infty} = 8.99\ \frac{MeV}{fm^{3}}$ by putting the energy
density of SQM and that of the nuclear matter equal to each other.

\subsection{Energy density calculation of strange quark matter in the presence of magnetic field}
We consider SQM composed of $u$, $d$ and $s$ quarks with spin up (+) and down (-).
We denote the number density of quark $i$ with spin up by $\rho_{i}^{(+)}$ and spin down by $\rho_{i}^{(-)}$.
We introduce the polarization parameter $\xi_{i}$ by
\begin{equation}\label{8}
\xi_{i} = \frac{\rho_{i}^{(+)} - \rho_{i}^{(-)}}{\rho_{i}},
\end{equation}
where $0 \leq \xi_{i} \leq 1$ and $\rho_{i} = \rho_{i}^{(+)} + \rho_{i}^{(-)}$. Under the conditions of
beta-equilibrium and charge neutrality, we get the following relation for the number density,
\begin{equation}\label{9}
\rho = \rho_{u} = \rho_{d} = \rho_{s},
\end{equation}
where $\rho$ is the total baryonic density of the system.

Within the MIT bag model, the total energy of SQM in the presence of  magnetic field ($B$) can be written as
\begin{equation}\label{11}
E_{tot} = E_K + \mathcal{B} + E_M,
\end{equation}
where $E_M$ is the contribution of magnetic energy, $\mathcal{B}$ is the bag constant (in this article,
we use a density dependent bag constant (Eq. (\ref{1})), and $E_K$ is the total kinetic energy of SQM.
The total kinetic energy of SQM is as follows,
\begin{equation}\label{121}
E_K=\sum_{i=u,d,s}E_{i},
\end{equation}
where $E_{i}$ is the kinetic energy of quark $i$,
\begin{equation}\label{12}
E_{i} = \sum_{p=\pm}\sum_{k^{(p)}} \sqrt{m_{i}^{2} c^{4} + \hbar^{2} k^{(p)^2} c^{2}}
\end{equation}
we ignore the masses of quark $u$ and $d$, while we assume $m_{s} = 150\ MeV$ for quark $s$.
After performing some algebra, supposing that $\xi_{s} = \xi_{u} = \xi_{d} = \xi$, we obtain the
following relation for the total kinetic energy density ($\varepsilon_K = \frac{E_K}{V}$) of SQM,
\begin{eqnarray}\label{13}
\varepsilon_K &=& \frac{3}{16 \pi^{2} \hbar^{3}} \sum_{p=\pm}[\frac{\hbar}{c^{2}} k_{F}^{(p)}
 E_{F}^{(p)} (2 \hbar^{2} k_{F}^{(p)^2} c^{2} + m_{s}^{2} c^{4}) - m_{s}^{4} c^{5}
ln (\frac{\hbar k_{F}^{(p)} + E_{F}^{(p)} /c}{m_{s} c})]\nonumber \\
&+& \frac{3 \hbar c \pi^{2/3}}{4} \rho^{4/3}[(1 + \xi)^{4/3} + (1 - \xi)^{4/3}]�
\end{eqnarray}
where
\begin{equation}\label{14}
k_{F}^{(\pm)} = (\pi^{2} \rho)^{1/3} (1 \pm \xi)^{1/3},
\end{equation}
and
\begin{equation}\label{15}
E_{F}^{(\pm)} = (\hbar^{2} k_{F}^{(\pm)^2} c^{2} + m_{s}^{2} c^{4})^{1/2}.
\end{equation}
For SQM, the contribution of magnetic energy is as $E_{M} = - M . B$. If we assume
the magnetic field is along the $z$ direction, the contribution of the magnetic energy of SQM is given by
\begin{equation}\label{16}
E_{M} = - \sum_{i=u,d,s} M_{z}^{(i)} B,
\end{equation}
where $M_{z}^{(i)}$ is the magnetization of the system corresponding to particle $i$ which is given by
\begin{equation}\label{17}
M_{z}^{(i)} = N_{i} \mu_{i} \xi_{i}.
\end{equation}
In the above equation, $N_{i}$ and $\mu_{i}$ are the number and magnetic moment of particle $i$, respectively.
By some simplification, the contribution of the magnetic energy density ($\varepsilon_M = \frac{E_M}{V}$) of SQM
can be obtained as follows
\begin{equation}\label{18}
\varepsilon_M = - \sum_{i=u,d,s} \rho_{i} \mu_{i} \xi_{i} B.
\end{equation}
Using the above equation and $\rho = \rho_{u} = \rho_{d} = \rho_{s}$ and with the assumption that
$\xi = \xi_{u} = \xi_{d} = \xi_{s}$, we have
\begin{equation}\label{19}
\varepsilon_M = - (\rho B \xi \mu_{s} + \rho B \xi \mu_{u} + \rho B \xi \mu_{d}).
\end{equation}
Now, we take advantage of numerical values of the magnetic moment for quarks ( Wong Samuel~\cite{rk40}) : \\
$ \mu_{s} = - 0.581\ \mu_{N} , \mu_{u} = 1.852\ \mu_{N}, \mu_{d} = - 0.972\ \mu_{N}$.\\
Using Eq. (\ref{19}) and above values, we conclude that
\begin{equation}\label{20}
\varepsilon_M = - 0.299 \rho \xi \mu_{N} B,
\end{equation}
where $\mu_{N} = 5.05 \times 10^{-27}\ (J/T)$ is nuclear magneton.

\subsection{The results of energy of strange quark matter in the presence of magnetic field}
We have calculated the properties of strange quark matter in the presence of magnetic field with the density
dependent bag constant (Eq. (\ref{1})). Our results are as follows.

\begin{figure}
\centering
\includegraphics[scale=0.28]{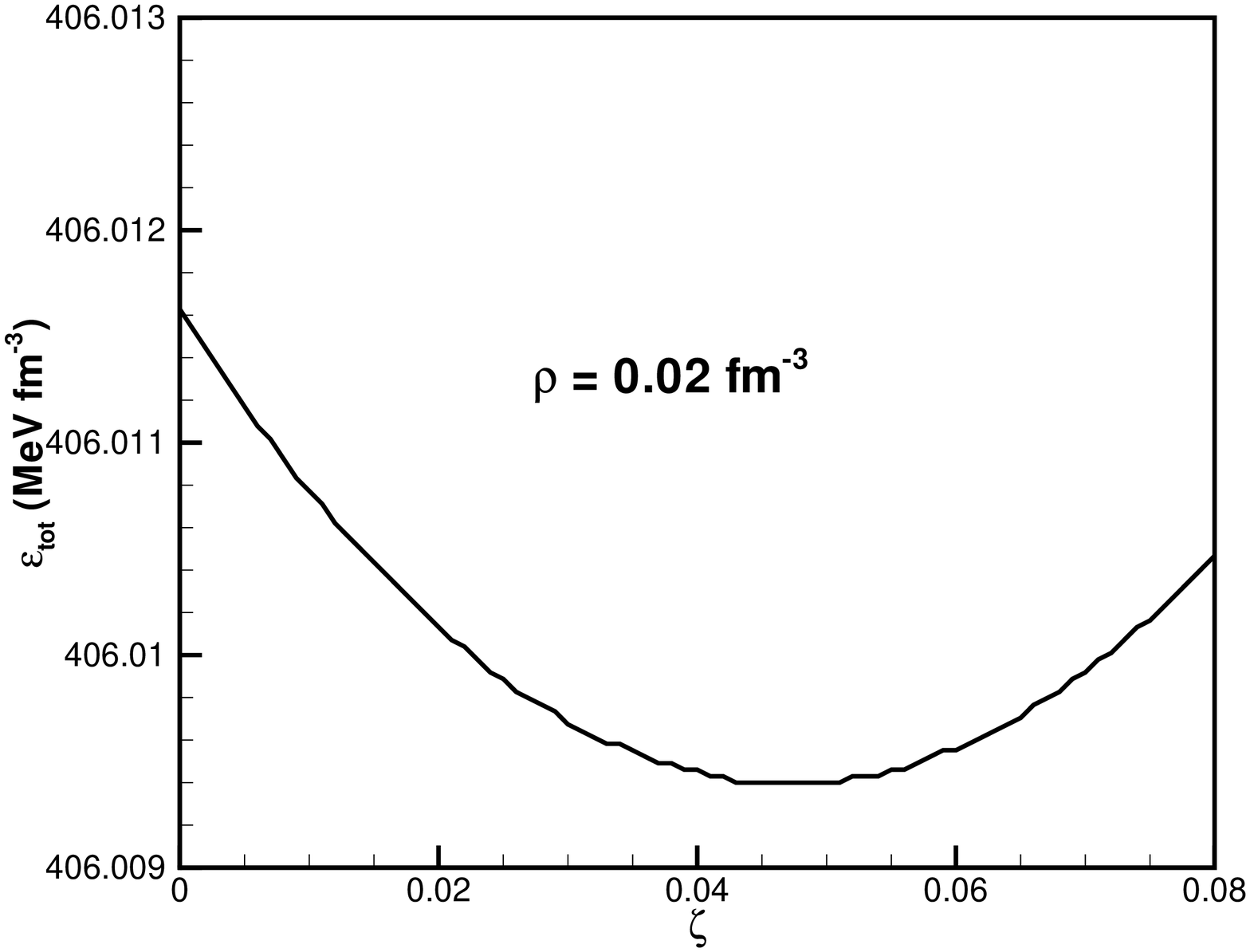}\quad
\includegraphics[scale=0.28]{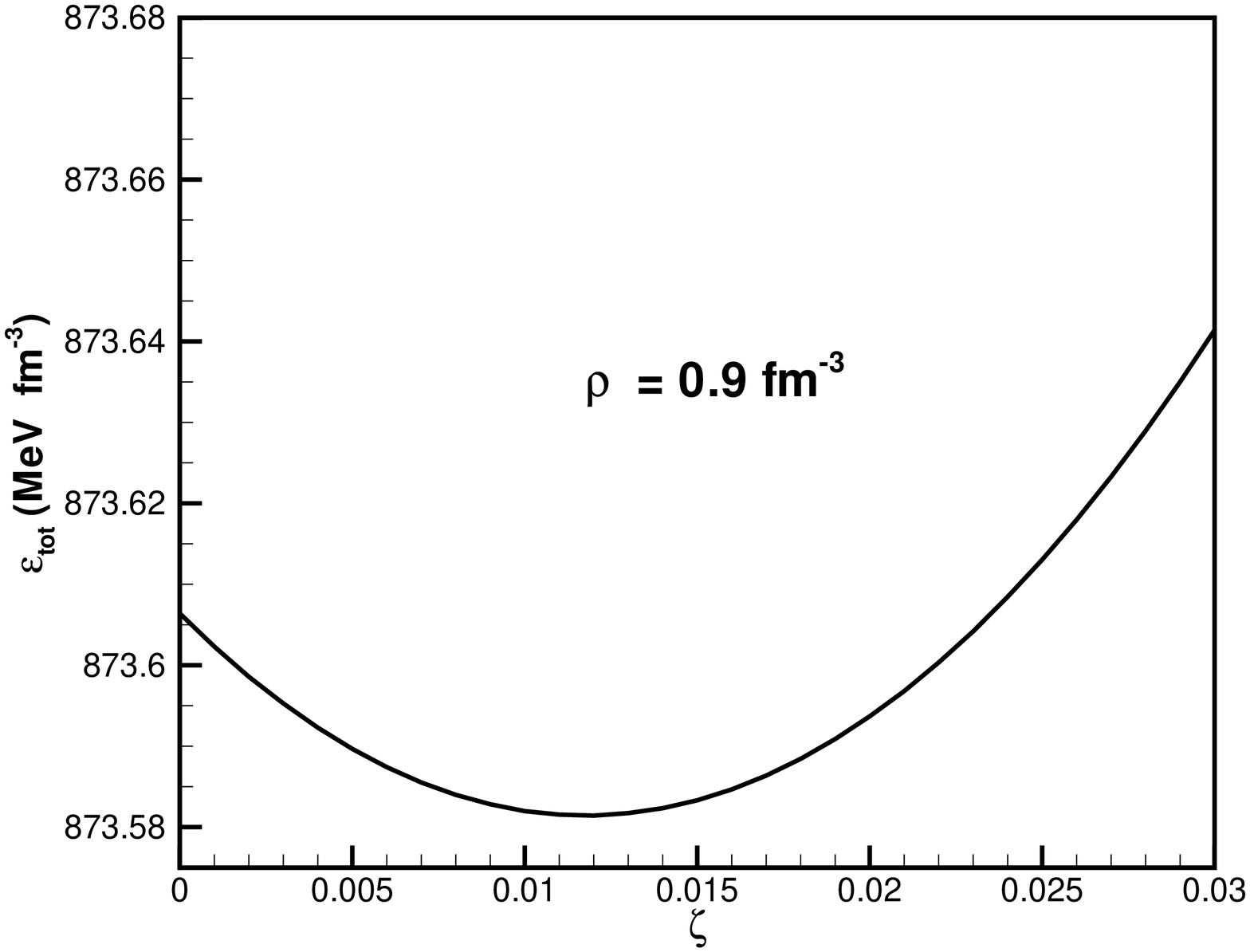}\quad
\includegraphics[scale=0.28]{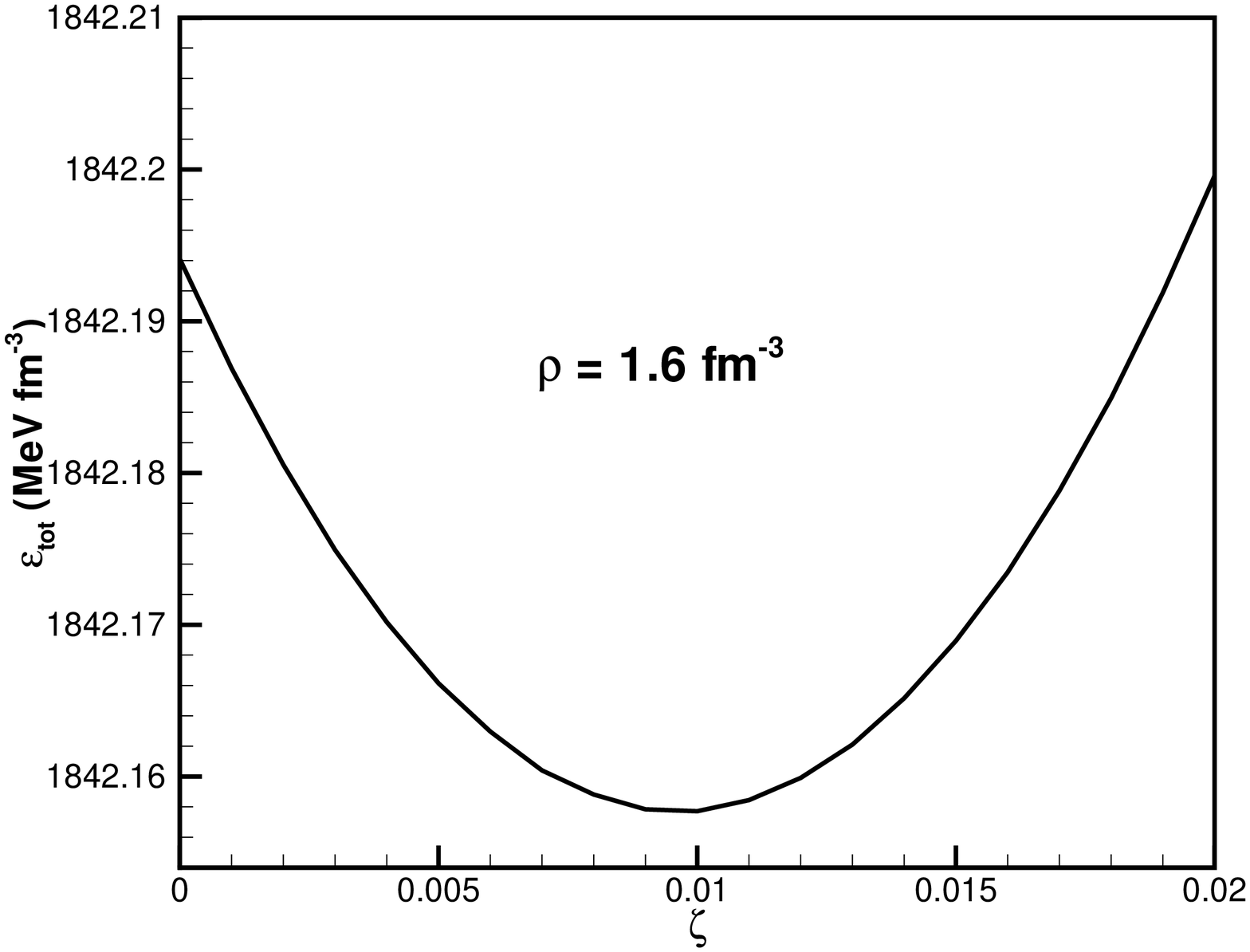}\quad
\includegraphics[scale=0.28]{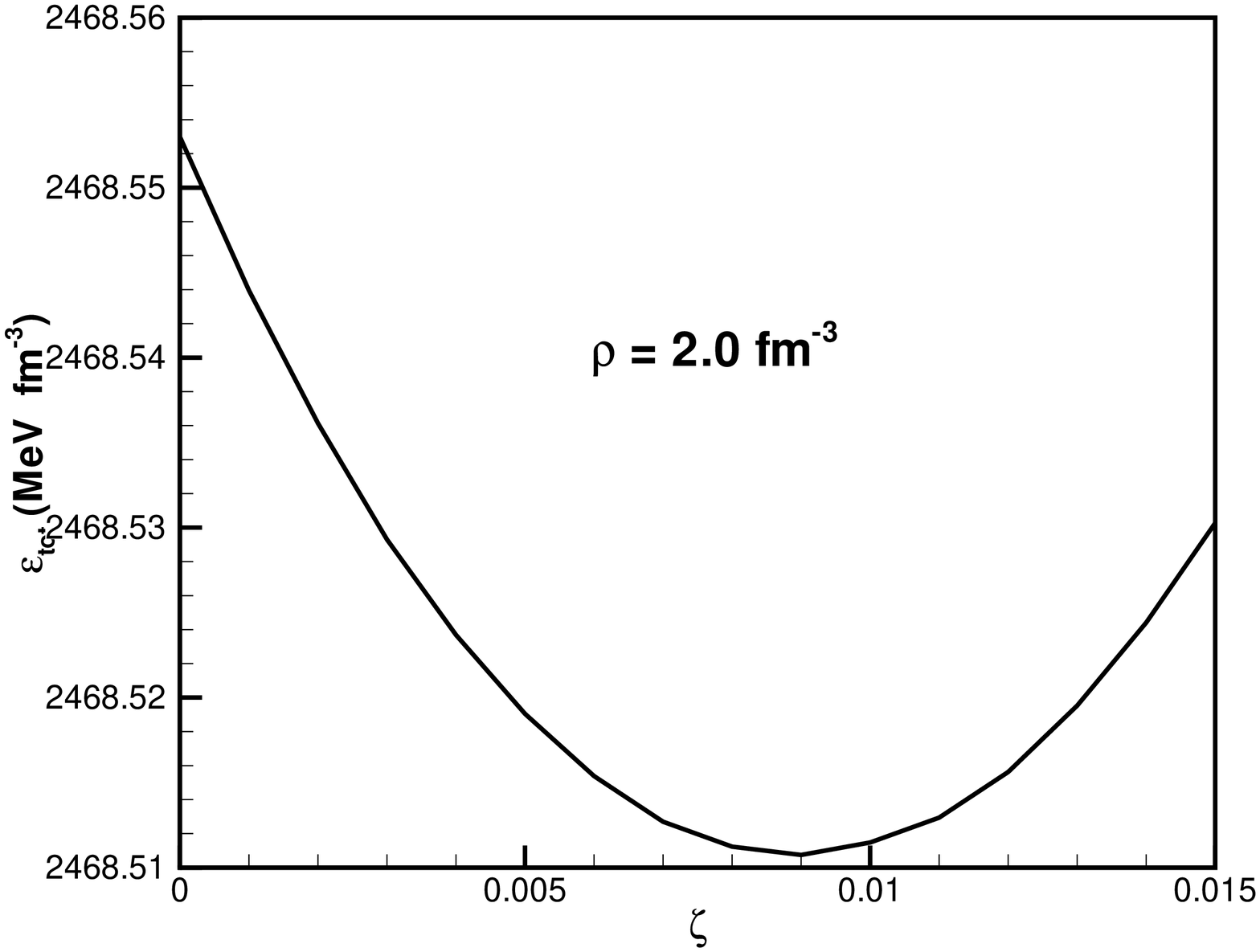}
\caption{Total energy density ($\varepsilon_{tot}$) as a function of the polarization parameter
($\xi$) for different densities ($\rho$).} \label{f1}
\end{figure}

Our results for the total energy density ($\varepsilon_{tot} = \frac{E_{tot}}{V}$) of strange quark matter
(SQM) in the presence of magnetic field
 have been plotted versus the polarization parameter in Fig. \ref{f1}  for various densities at
$B = 5 \times 10^{18}\ G$.
We see that there is a minimum point in the energy curve for each density which shows a meta-stable state for this system.
As the density increases, the minimum point of energy shifts to the lower values of the polarization, and finally
it disappears at high densities in which the system becomes nearly unpolarized.

\begin{figure}
\centering
\includegraphics[width=\textwidth, angle=0]{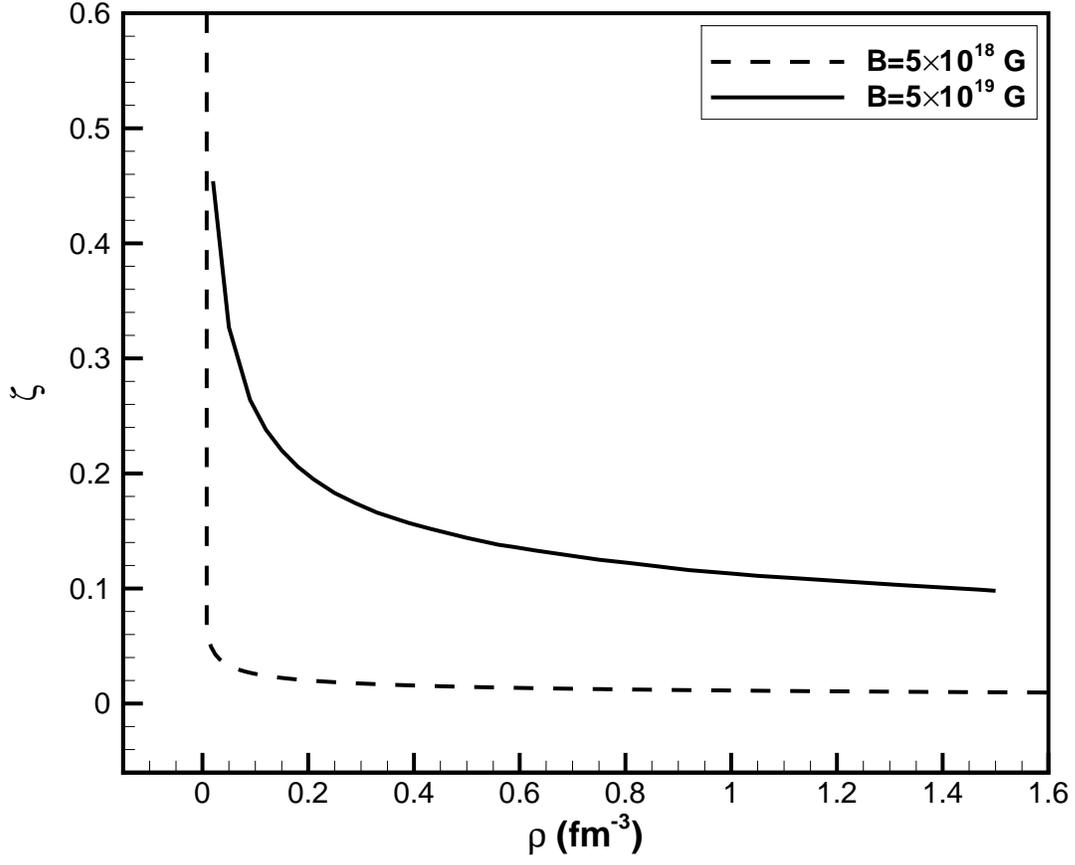}
\caption{Polarization parameter ($\xi$) versus density ($\rho$) for $B = 5 \times 10^{18}$ and
$5 \times 10^{19}\ G$.} \label{f2}
\end{figure}

\begin{figure}
\centering
\includegraphics[width=\textwidth, angle=0]{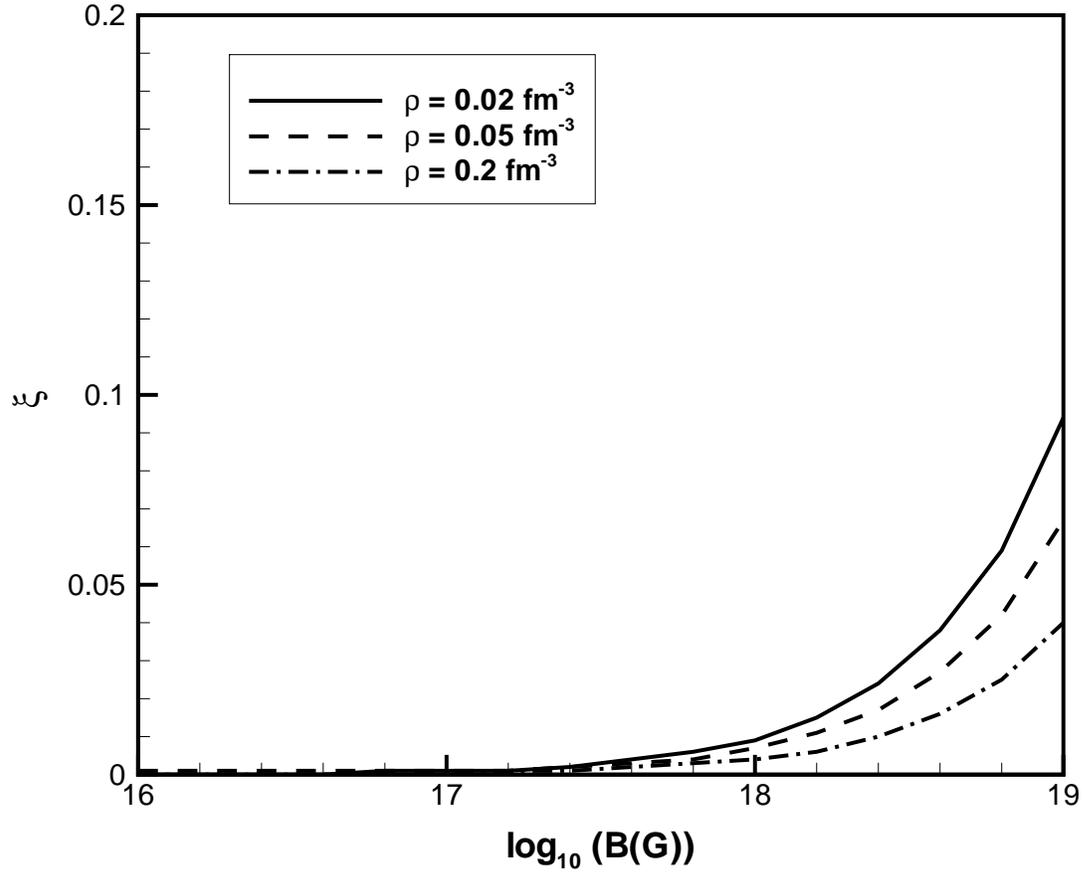}
\caption{Polarization parameter ($\xi$) corresponding to the minimum points of energy density versus the
magnetic field ($B$) for different values of density ($\rho$). } \label{f3}
\end{figure}

In Fig. \ref{f2}, we have plotted the polarization parameter corresponding to the minimum point of  energy versus density for
two magnetic fields $B = 5 \times 10^{18}\ G$ and $B = 5 \times 10^{19}\ G$. We can see that the value of $\xi$
decreases by increasing the density, and it becomes nearly zero at high densities.
We have also drown the polarization parameter as a function of the magnetic field at different densities in Fig. \ref{f3}.
As this figure shows, the polarization parameter increases by increasing the magnetic field for all densities.

\begin{figure}
\centering
\includegraphics[width=\textwidth, angle=0]{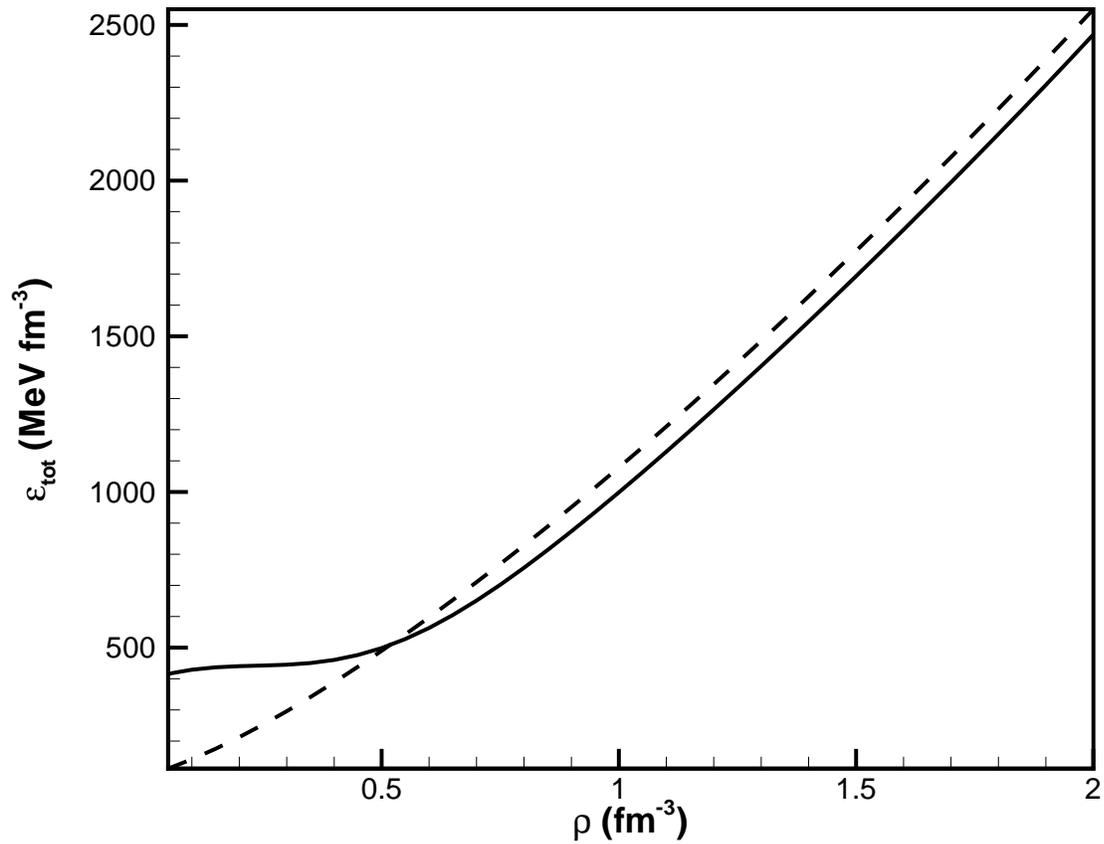}
\caption{Total energy density versus density ($\rho$) calculated by density dependent bag
constant (full curve) at $B = 5 \times 10^{18}\ G$. The results for $\mathcal{B} = 90\ \frac{MeV}{fm^{3}}$
(dashed curve) at $B = 5 \times 10^{18}\ G$ have been also given for comparison.} \label{f4}
\end{figure}

For the strange quark matter, our results for the total energy density at $B = 5 \times 10^{18}\ G$, which calculated with the
density dependent bag constant, have been shown as a function of density in Fig. \ref{f4}.
The results for $\mathcal{B} = 90\ \frac{MeV}{fm^{3}}$
at $B = 5 \times 10^{18}\ G$ (Bordbar \& Peyvand~\cite{rk20}) are also given for comparison.
It can be observed that the total energy density
has an increasing rate by increasing the density. Also, it can be found that for $\rho$ greater (lower) than about $0.6\ fm^{-3}$,
the energy of SQM with the density dependent bag constant is lower (greater) than that with the fixed bag constant.
From Fig. \ref{f4}, it is seen that for $\rho<0.6\ fm^{-3}$, the increasing of energy has a slow slope, whereas
for $\rho>0.6\ fm^{-3}$ this increasing is accomplished with a  more quick slope.

\begin{figure}
\centering
\includegraphics[width=\textwidth, angle=0]{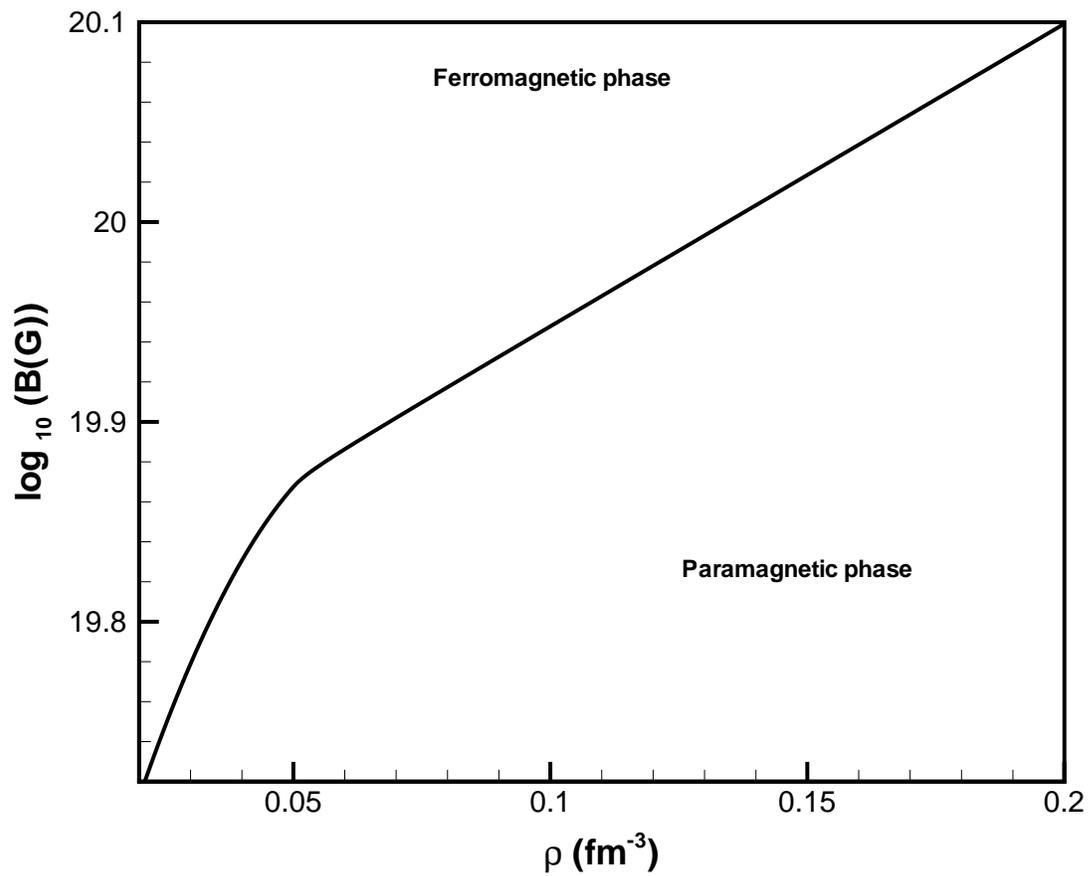}
\caption{Phase diagram for the strange quark matter in the presence of a strong magnetic field. } \label{f5}
\end{figure}

Fig. \ref{f5} shows the phase diagram for the strange quark matter. We can see that as the density
increases, the magnetic field grows monotonically.
It explicitly means that at higher densities, the ferromagnetic phase transition occurs
at higher values of the magnetic field.

\subsection{The equation of state for strange quark matter in the presence of magnetic field}
\begin{figure}
\centering
\includegraphics[scale=0.5]{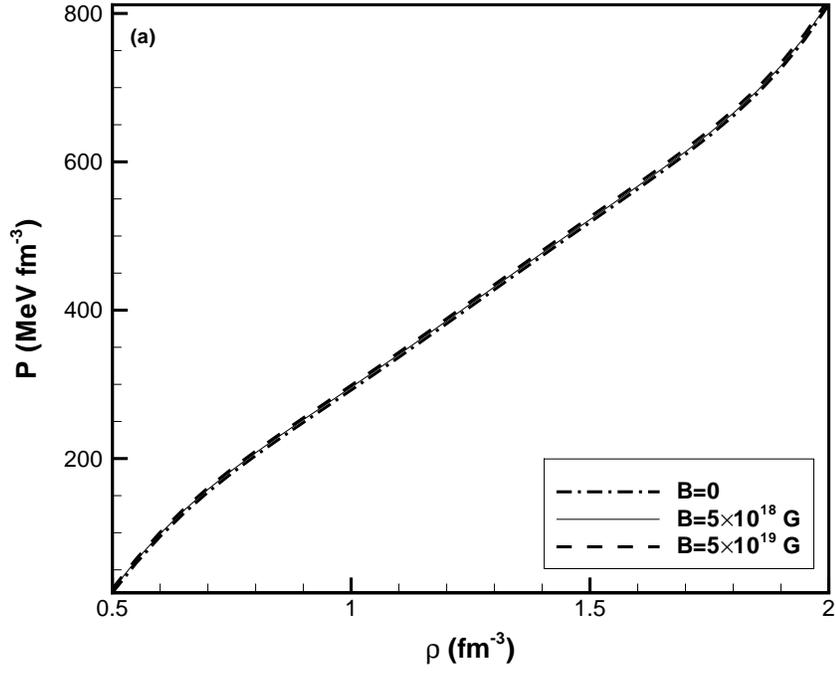}\quad
\includegraphics[scale=0.46]{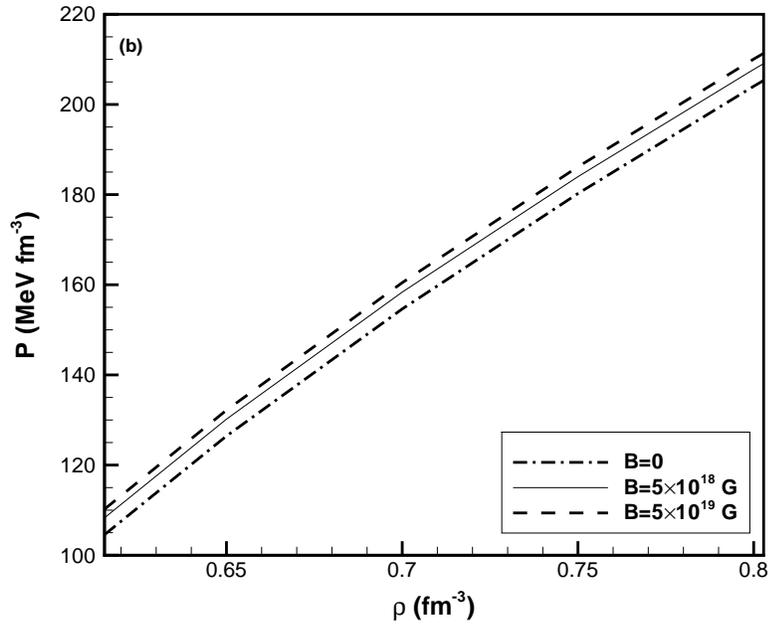}
\caption{The equation of state of SQM  at $B = 0$, $\ 5 \times 10^{18}\ $ and $5 \times 10^{19}\ G$. } \label{f6}
\end{figure}

\begin{figure}
\centering
\includegraphics[width=\textwidth, angle=0]{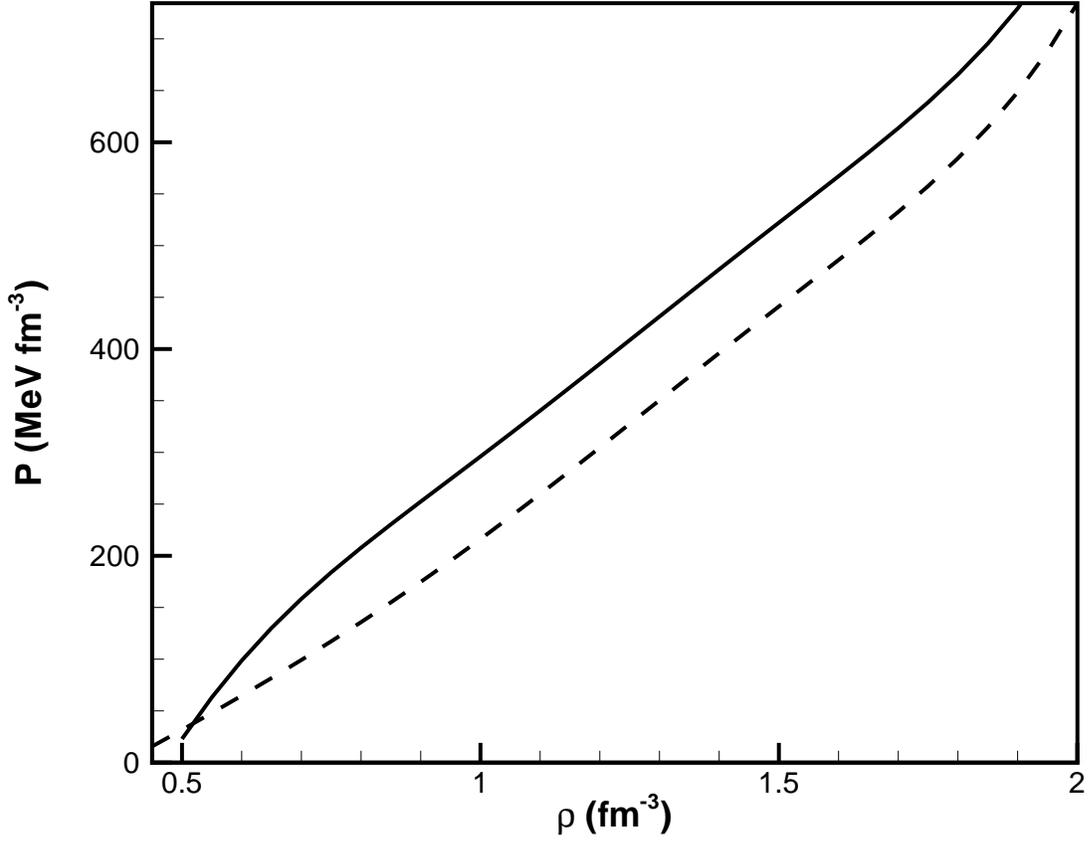}
\caption{The equation of state of SQM in the case of density dependent bag constant (full curve) at
$B = 5 \times 10^{18}\ G$. The results for the case of fixed bag constant ($\mathcal{B} = 90\ \frac{MeV}{fm^{3}}$)
(dashed curve) at $B = 5 \times 10^{18}\ G$ have been also given for comparison.} \label{f7}
\end{figure}

In this section, we calculate the equation of state (EoS) of strange quark matter (SQM) in the presence of magnetic field
with density dependent bag constant.
Generally, we can calculate EoS using the following relation,
\begin{equation}\label{22}
P(\rho) = \rho \frac{\partial \varepsilon_{tot}}{\partial \rho} - \varepsilon_{tot},
\end{equation}
where $P$ is the pressure and $\varepsilon_{tot}$ is the energy density, which in the presence of  magnetic
field, is obtained from  Eq. (\ref{11}).
In Fig. \ref{f6}, we have compared our results for EoS of SQM at different magnetic fields.
This shows that for all magnetic fields, by increasing the density, pressure has an increasing rate. Also, we can see
that with increasing magnetic field, the pressure increases.
In Fig. \ref{f7}, we have drawn EOS of SQM by density dependent bag constant at
$B = 5 \times 10^{18}\ G$. The results for $\mathcal{B} = 90\ \frac{MeV}{fm^{3}}$ at $B = 5 \times 10^{18}\ G$
(Bordbar \& Peyvand~\cite{rk20}) are also given
for comparison. This figure indicates that for $\rho$ greater than about $0.52\ fm^{-3}$,
when  the bag constant is density dependent, the pressure of SQM
is greater than that of the density independent case.

\section{Structure of strange quark star}
Quark stars are relativistic objects, therefore we used the general
relativity for calculation of their structures. Since most of the massive general relativistic objects have some forms
of rotation (very rapid in the case of pulsars). In this calculations, we are interested in the
investigation of the strong magnetic field effects on the structure of a static strange quark star. Using the the equation of state (EoS) of strange quark matter (SQM), we can obtain
the structure of these stars by numerically integrating the
general relativistic equations of hydrostatic equilibrium, the Tolman-Oppenheimer-Volkoff
(TOV) equation (Shapiro \& Teukolski~\cite{rk9}),
\begin{eqnarray}\label{25}
\frac{dP}{dr} &=& - \frac{G  (\frac{\varepsilon (r) m(r)}{r^{2}})(1 + \frac{P(r)}{c^{2} \varepsilon(r)})
(1 + \frac{4 \pi r^{3} P(r)}{m(r) c^{2}} )}{ (1 - \frac{2 G m(r)}{c^{2}})}\nonumber\\
\frac{dm}{dr} &=& 4 \pi r^{2} \varepsilon(r).
\end{eqnarray}
In the above equations, $P$ is pressure and $\varepsilon(r)$ is the energy
density, $G$ is the gravitational constant and $m(r)$ is the mass inside radius $r$
which is calculated as follows
\begin{eqnarray}\label{26}
m(r) =  \int_{0}^{r} 4 \pi r'^{2} \varepsilon(r') d r'.
\end{eqnarray}
Now, by selecting a central energy density $\varepsilon_{c}$, under the boundary conditions
$P (0) = P_{c}$ and $m (0) = 0$,
we integrate TOV equations outwards to a radius $r = R$ at which $P$ vanishes
(Shapiro \& Teukolski~\cite{rk9}).

In this section, we calculate the structure of the SQS with density
dependent bag constant in the presence of a magnetic field. We
should  note that a strong magnetic field changes the spherical
symmetry of the system and for magnetic fields less than $10^{19}\
G$, this effect is ignorable (Felipe \& Martinez~\cite{rk42};
Gonzalez Felipe et al.~\cite{rk43}). Considering the anisotropy of
the strange quark matter pressures in the presence of magnetic
field, it has been shown that for vanishing AMM (anomalous magnetic
moments), the perpendicular component of the pressure $P_{\perp}$
goes to zero at about $2\times10^{19}G$ (Gonzalez Felipe et
al.~\cite{rk56}). Thus in the case of SQM, for $B<10^{19}G$ the
anisotropy in the pressures is relatively small, i.e,
$P_{\perp}=P_{\parallel}$.

\begin{figure}
\centering
\includegraphics[width=\textwidth, angle=0]{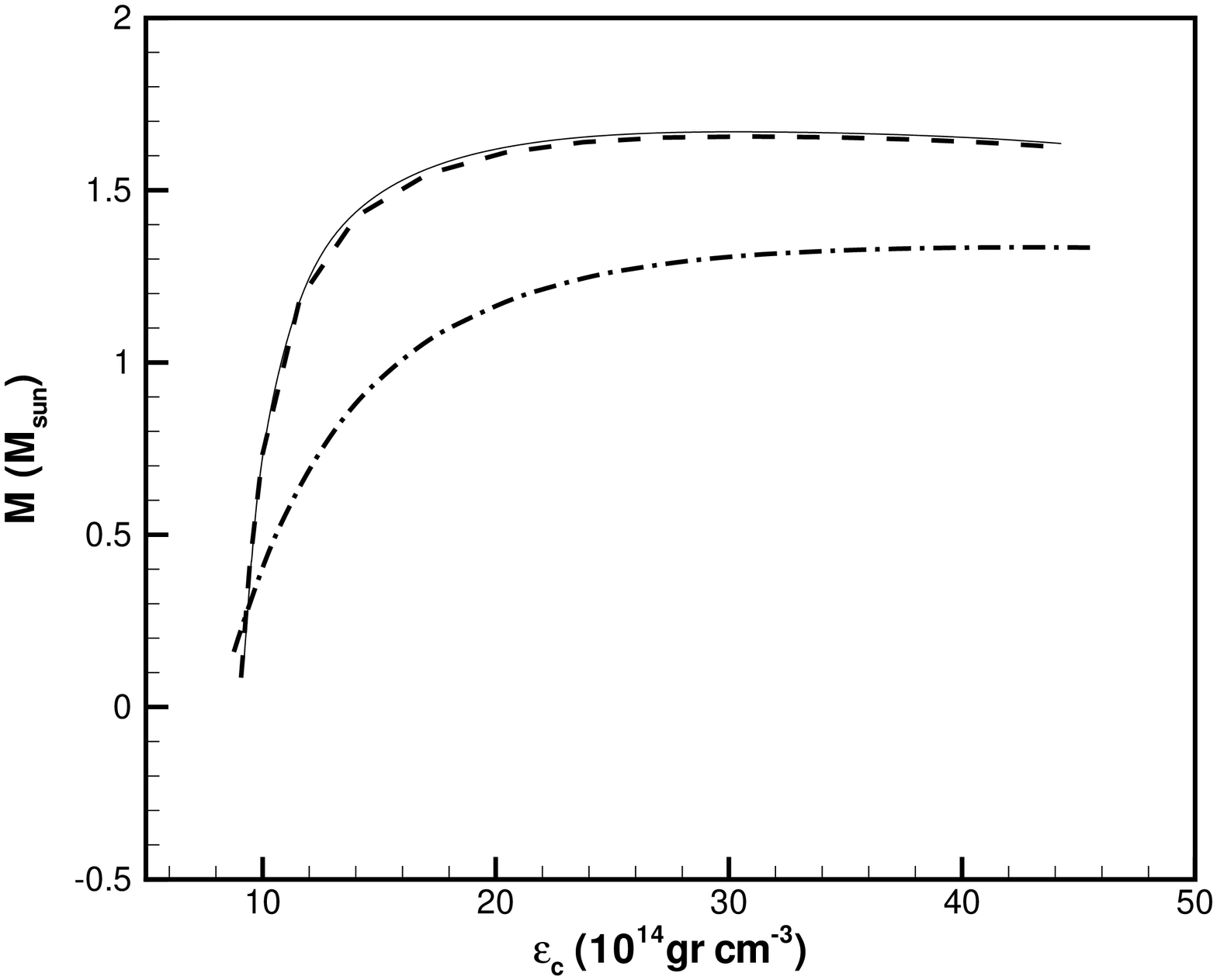}
\caption{Gravitational mass versus the central energy density ($\varepsilon_{c}$) at $B = 0$
(full curve) and $B = 5 \times 10^{18}\ G$ (dashed curve). The results for $\mathcal{B} = 90\ \frac{MeV}{fm^{3}}$
(dashed dotted curve) at $B = 5 \times 10^{18}\ G$ have been also given for comparison.} \label{f8}
\end{figure}

In Fig. \ref{f8}, we have drawn the gravitational mass versus the central density ($\varepsilon_{c}$) for
an SQS in the magnetic fields $B = 0$ and $5 \times 10^{18}\ G$.
We see that as the central density increases,
the gravitational mass of an SQS increases, and finally it reaches a limiting value which
is called maximum gravitational mass. Fig. \ref{f8} shows that by presenting the magnetic field,
the gravitational mass decreases. The results for $\mathcal{B} = 90\ \frac{MeV}{fm^{3}}$
at $B = 5 \times 10^{18}\ G$ (Bordbar \& Peyvand~\cite{rk20}) are also given in Fig. \ref{f8}
for the sake of comparison. This indicates that the density dependence of bag constant leads to
substantially higher values for the gravitational mass of SQS. With the density dependent bag constant,
we have found that the maximum gravitational mass of SQS is about $1.62\ M_\odot$,
while with the fixed bag constant, it is about $1.33\ M_\odot$.

\begin{figure}
\centering
\includegraphics[width=\textwidth, angle=0]{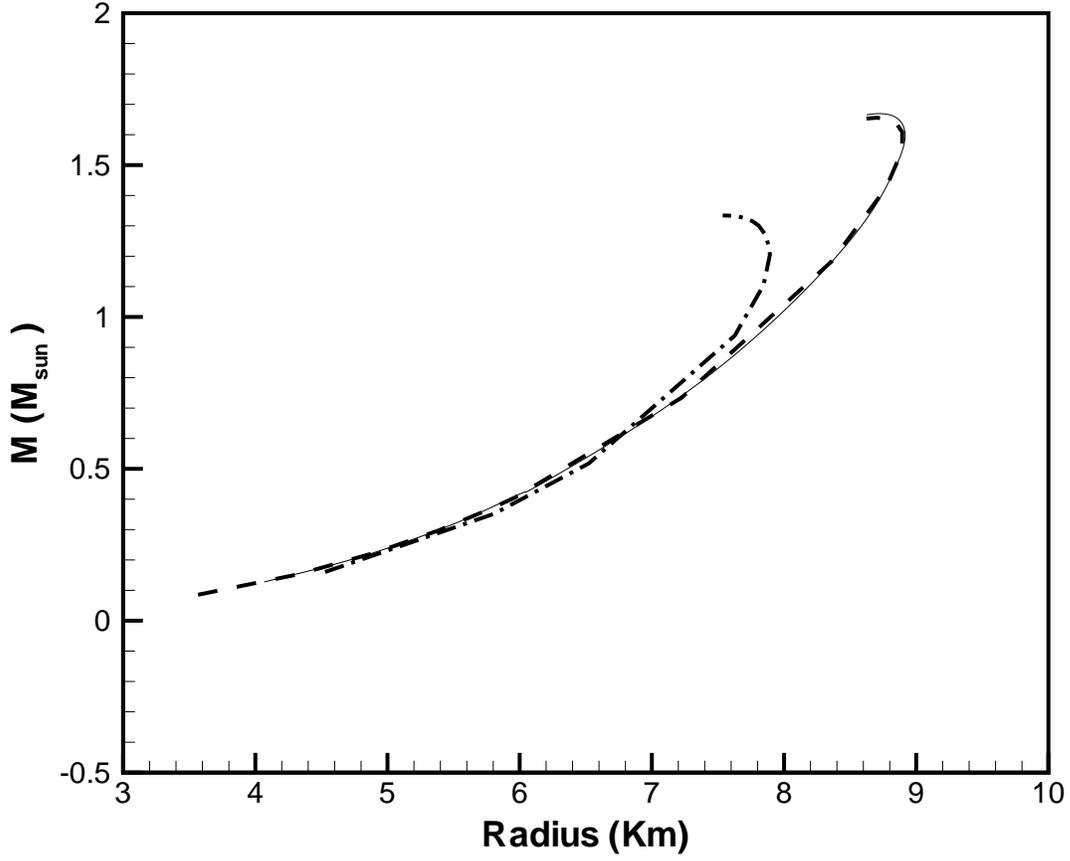}
\caption{The gravitational mass versus radius at $B = 0$ (full curve)
and $B = 5 \times 10^{18}\ G$ (dashed curve). The results for $\mathcal{B} = 90\ \frac{MeV}{fm^{3}}$ (dashed dotted curve)
at $B = 5 \times 10^{18}\ G$ have been also given for comparison.} \label{f9}
\end{figure}

We have plotted the gravitational mass of SQS as a function of the radius (mass-radius relation)
for the  magnetic fields $B = 0$ and $5 \times 10^{18}\ G$ in Fig. \ref{f9}.
It is seen that for all cases of SQS, the gravitational mass increases by increasing the radius.
In Fig. \ref{f9}, we have also compared our results for the density dependent case of bag constant
with those of density independent case. We can see that for the case of fixed bag constant,
the increasing rate of gravitational mass versus radius is higher than that of density dependent case.
However, it will be more constructive to consider the effects of rotation on the properties of the star which is beyond our present investigation. Some authors have shown that considering the rotation of the star
leads to the larger maximum mass for strange quark stars (Shen et al.~\cite{rk57}).

%%%%%%%%%%%%%%%%%%%%%%%%%%%%%%%%%%%%%%%%%%%%%%%%%%%%%%%%%%%%%%%%
\section{Summary and conclusions}\label{V}
We have investigated a cold static strange quark star in the presence of a strong magnetic field.
For this purpose, some of the bulk properties of the strange quark matter such as the energy density and
equation of state have been computed using the MIT bag model with the density dependent bag constant.
Calculations of the energy for different magnetic polarization  in the presence of a magnetic field
demonstrated that as the density increases,
the minimum point of energy shifts to the lower values of the polarization. We have shown that the value
of the polarization parameter decreases by increasing the density, and it also increases by increasing the
magnetic field. Our results at $B = 5 \times 10^{18}\ G$ show that
for both cases of the density dependent bag constant and fixed bag constant, the
total energy density have an increasing rate by increasing the density. We have shown that there is a ferromagnetic
phase transition at high magnetic fields. Our computations indicate that the pressure increases by increasing the density
and magnetic field. In this work, we have also studied the structure properties of the strange quark stars.
Our results show that the gravitational mass of the strange quark star increases by increasing
the central energy density. It was shown that this gravitational mass reaches a limiting value at higher values
of the central energy density. We have shown that the maximum mass of the strange quark star  reduces by presenting the magnetic
field. Finally, a comparison has been also made between the results of density dependent bag constant and those of fixed bag constant.
Our calculation with the density dependent bag constant shows a higher maximum mass with respect to that of fixed bag constant.

One of the possible astrophysical implications of our results is
calculation of the surface redshift $(z_{s})$ of SQS. This parameter
is of special interest in astrophysics and can be obtained from the
mass and radius of the star using the following relation (Camenzind
~\cite{rk15}),
\begin{equation}
z_{s}=(1-\frac{2GM}{Rc^{2}})^{-\frac{1}{2}}-1.
\end{equation}
Our results corresponding to the maximum mass and radius of SQS
calculated by density dependent bag constant lead to $ z_{s}=0.534 \
ms^{-1}$ for the magnetic field $B=5\times10^{18}G$.
%%%%%%%%%%%%%%%%%%%%%%%%%%%%%%%%%%%%%%%%%%%%%%%%%%%%%%%%%%%%%%%%%%%
\section*{Acknowledgements}
This work has been supported by Research Institute for Astronomy
and Astrophysics of Maragha. We wish to thank Shiraz University
Research Council.

%%%%%%%%%%%%%%%%%%%%%%%%%%%%%%%%%%%%%%%%%%%%%%%%%%%%%%%%%%%%%%%%%%

%%%%%%%%%%%%%%%%%%%%%%%%%%%%%%%%%%%%%%%%%%%%%%%%%%%%%%%%%%%%%
%%%%%%%%%%%%%%%%%%%%%%%%%%%%%%%%%%%%%%%%%%%%%%%%%%%%%%%%%%%%%%%%%%%%%%%%%%%%%%%%%%%%%%%%%%%%%%%%%%%%%%
\newpage
%%%%%%%%%%%%%%%%%%%%%%%%%%%%%%%%%%%%%%%%%%%%%%%%%%%%%%%%%%%%%%%%%%%%%%%%%%%%%%%%%%%%%%%%%%%%%%%%%%%%%%
\newpage
%%%%%%%%%%%%%%%%%%%%%%%%%%%%%%%%%%%%%%%%%%%%%%%%%%%%%%%%%%%%%%%%%%%%%%%%%%%%%%%%%%%%%%%%%%%%%%%%%%%%%%
\newpage
%%%%%%%%%%%%%%%%%%%%%%%%%%%%%%%%%%%%%%%%%%%%%%%%%%%%%%%%%%%%%%%%%%%%%%%%%%%%%%%%%%%%%%%%%%%%%%%%%%%%%%
\newpage
%%%%%%%%%%%%%%%%%%%%%%%%%%%%%%%%%%%%%%%%%%%%%%%%%%%%%%%%%%%%%%%%%%%%%%%%%%%%%%%%%%%%%%%%%%%%%%%%%%%%%%
\newpage
%%%%%%%%%%%%%%%%%%%%%%%%%%%%%%%%%%%%%%%%%%%%%%%%%%%%%%%%%%%%%%%%%%%%%%%%%%%%%%%%%%%%%%%%%%%%%%%%%%%%%%
\newpage
%%%%%%%%%%%%%%%%%%%%%%%%%%%%%%%%%%%%%%%%%%%%%%%%%%%%%%%%%%%%%%%%%%%%%%%%%%%%%%%%%%%%%%%%%%%%%%%%%%%%%%
\newpage
%%%%%%%%%%%%%%%%%%%%%%%%%%%%%%%%%%%%%%%%%%%%%%%%%%%%%%%%%%%%%%%%%%%%%%%%%%%%%%%%%%%%%%%%%%%%%%%%%%%%%%
\newpage
%%%%%%%%%%%%%%%%%%%%%%%%%%%%%%%%%%%%%%%%%%%%%%%%%%%%%%%%%%%%%%%%%%%%%%%%%%%%%%%%%%%%%%%%%%%%%%%%%%%%%%
\newpage
\end{document}